\documentclass[twocolumn,floatfix,prl,showpacs,superscriptaddress]{revtex4}
\usepackage{graphicx,color}

\begin{document}
\title{
Magnetic excitations in the spin-1/2 triangular-lattice antiferromagnet Cs$_2$CuBr$_4$}
\author{S.~A. Zvyagin}

\affiliation{Dresden High Magnetic Field Laboratory (HLD-EMFL), Helmholtz-Zentrum Dresden-Rossendorf, 01328 Dresden,
Germany}

\author{M.~Ozerov}
\thanks{Present address: FELIX Laboratory, Radboud University  Nijmegen, 6525 ED Nijmegen, The Netherlands}
\affiliation{Dresden High Magnetic Field Laboratory (HLD-EMFL), Helmholtz-Zentrum Dresden-Rossendorf, 01328 Dresden, Germany}

\author{D. Kamenskyi}
\thanks{Present address: High Field Magnet Laboratory (HFML-EMFL), Radboud University  Nijmegen, 6525 ED Nijmegen, The Netherlands}
\affiliation{Dresden High Magnetic Field Laboratory (HLD-EMFL), Helmholtz-Zentrum Dresden-Rossendorf, 01328 Dresden, Germany}

\author{J. Wosnitza}
\affiliation{Dresden High Magnetic Field Laboratory (HLD-EMFL), Helmholtz-Zentrum Dresden-Rossendorf, 01328 Dresden,
Germany}

\affiliation{Institut f\"{u}r Festk\"{o}rperphysik, TU Dresden, 01062 Dresden, Germany}

\author{J. Krzystek}
\affiliation{National High Magnetic Field Laboratory, Florida
State University, Tallahassee, FL 32310, USA}

\author{D.~Yoshizawa}
\affiliation{Center for Advanced High Magnetic Field Science (AHMF), Graduate School of Science, Osaka University, Toyonaka, Osaka 560-0043, Japan}

\author{M. Hagiwara}
\affiliation{Center for Advanced High Magnetic Field Science (AHMF), Graduate School of Science, Osaka University, Toyonaka, Osaka 560-0043, Japan}

\author{Rongwei Hu}
\altaffiliation[Present address: ]
{Rutgers Center for Emergent Materials and
Department of Physics and Astronomy, Rutgers University, Piscataway, NJ 08854, USA}
\affiliation{Condensed Matter Physics and Materials Science Department, Brookhaven National Laboratory, Upton,
NY 11973, USA}

\author{Hyejin Ryu}
\affiliation{Condensed Matter Physics and Materials Science Department, Brookhaven National Laboratory, Upton,
NY 11973, USA}
\affiliation{Department of Physics and Astronomy, Stony Brook University, Stony Brook, New York 11794-3800, USA}

\author{C. Petrovic}
\affiliation{Condensed Matter Physics and Materials Science Department, Brookhaven National Laboratory, Upton,
NY 11973, USA}
\affiliation{Department of Physics and Astronomy, Stony Brook University, Stony Brook, New York 11794-3800, USA}

\author{M.~E. Zhitomirsky}
\affiliation{Service de Physique Statistique, Magn\'etisme et Supraconductivit\'e,
UMR-E9001 CEA-INAC/UJF,  38054 Grenoble Cedex 9, France}

\begin{abstract}
We report on  high-field electron spin resonance (ESR) studies of magnetic excitations in the spin-1/2 triangular-lattice antiferromagnet Cs$_2$CuBr$_4$. Frequency-field diagrams of ESR excitations are measured for different orientations of magnetic fields up to 25 T.
We show that the substantial  zero-field energy gap, $\Delta\approx9.5$ K, observed  in the low-temperature excitation spectrum of Cs$_2$CuBr$_4$  [Zvyagin $et~al.$, Phys. Rev. Lett. 112, 077206 (2014)], is present well above  $T_N$.    Noticeably, the transition into the long-range magnetically ordered phase  does not significantly affect the size of the gap, suggesting that even  below   $T_N$ the high-energy spin dynamics in Cs$_2$CuBr$_4$ is  determined by short-range-order spin correlations.  The experimental data are compared with results of model spin-wave-theory calculations for  spin-1/2 triangle-lattice antiferromagnet.
\end{abstract}

\maketitle

Spin-1/2 Heisenberg antiferromagnets (AFs) on triangular lattices form
an important class of  low-dimensional (low-D) spin systems to probe  effects of quantum fluctuations,
magnetic order, and frustrations. In 1973,
developing the idea of  the ``resonating valence bond''  ground state,
P.~W.~Anderson proposed  that quantum fluctuations can be sufficiently strong to destroy the classical $120^{\circ}$ order
in such a system  \cite{Anderson}. As consequence, a two-dimensional   spin-liquid phase may be realized.
This phase can be regarded as a 2D fluid of
resonating spin-singlet pairs,
with the excitation spectrum formed by fractionalized mobile quasiparticles.
Subsequent numerical studies have, however, confirmed the presence of the semiclassical $120^{\circ}$ magnetic
ordering albeit with strongly reduced ordered moments \cite{Bernu,Capriotti,Zheng,White}. In recent years, the topic of spin-1/2 Heisenberg AFs on a triangular lattice has received particular
attention due to the  rich phase diagram,  whose details are still actively debated   \cite{Starykh,Gham,Heid,Weng,Weich,Reuther}.

Cs$_2$CuCl$_4$  and Cs$_2$CuBr$_4$  are two  prominent members of this family of frustrated  spin systems. Inelastic neutron-scattering experiments in Cs$_2$CuCl$_4$  revealed  the presence of a highly dispersive continuum of excited states \cite{Coldea_2D_SL}. These excitations were    initially identified as spinons  in the 2D frustrated spin liquid. Later on, the data have been re-interpreted in the frame of the 1D spin-liquid scenario \cite{Kohno,Balents} with  interchain bound spinon excitations  (triplons) as a signature of the quasi-1D nature of  magnetic correlations in this material \cite{Spin_Reduction,rem5}.  Electron spin resonance (ESR) studies  provided   additional  support for the proposed quasi-1D Heisenberg AF chain model with the uniform Dzyaloshinskii-Moriya  (DM) interaction,  opening an energy gap, $\Delta = 14$ GHz,  at the $\Gamma$ point   \cite{Povarov}.

\begin{figure}[h]
\begin{center}
\vspace{0cm}
\includegraphics[width=0.5\textwidth]{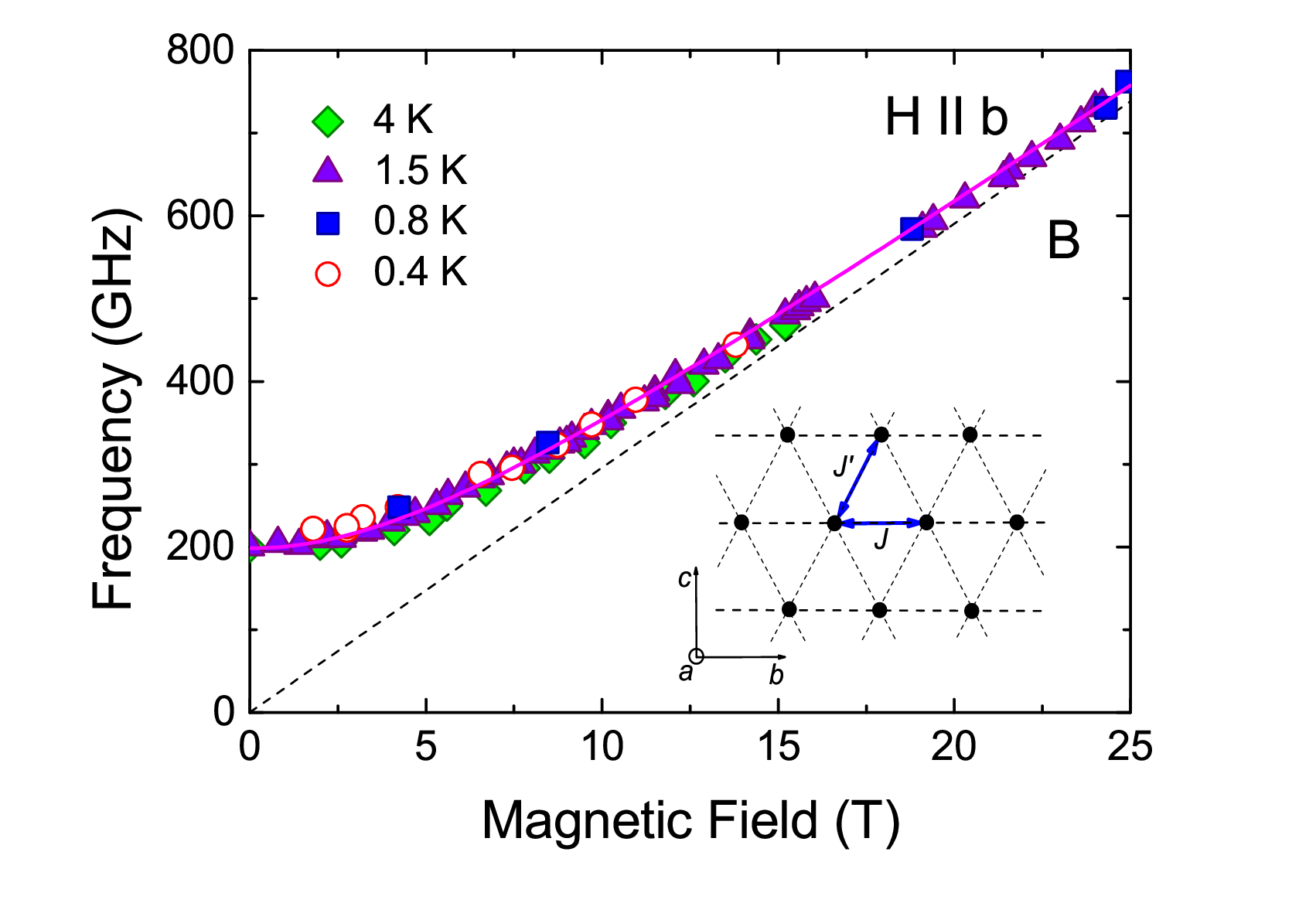}
\vspace{-0.8cm}
\caption{\label{fig:CCB_FFD_b} (color online) Frequency-field diagram of ESR excitations obtained for Cs$_{2}$CuBr$_{4}$
with magnetic field applied along the $b$ axis at different temperatures (mode B).
Fit  results for 1.5 K using Eq. (\ref{CCB_FFD_b})  are shown by the solid line.  The frequency-field dependence  of the paramagnetic resonance with $g_b=2.11$  is shown by the dashed line. The inset shows a schematic picture of exchange paths in the $bc$ plane of Cs$_{2}$CuBr$_{4}$. }
\end{center}
\end{figure}

Cs$_{2}$CuBr$_{4}$  realizes a distorted triangular lattice with orthorhombic crystal structure,  space group $Pnma$,  and
the room-temperature lattice  parameters $a = 10.195$~{\AA}, $b = 7.965$~{\AA}, and
$c = 12.936$~{\AA} ($Z=4$) \cite{Morozin}.
Compared  to  $J'/J \simeq  0.3$ \cite{CCC-par,CCB_Zvyagin} for Cs$_2$CuCl$_4$, the $J'/J$ ratio for Cs$_2$CuBr$_4$
is  somewhat larger, $J'/J \simeq 0.41$ \cite{CCB_Zvyagin}, that places  this compound further away from the decoupled
AF chain limit  and  makes it  more frustrated.  This difference  is thought to be related to the 1/3 magnetization plateau  and the cascade of field-induced phase transitions,  observed in Cs$_2$CuBr$_4$  \cite{Radu_BEC,Fortune}. In spite of intensive theoretical and experimental efforts, very  little is known about the spin dynamics in  Cs$_2$CuBr$_4$.  Inelastic neutron-scattering studies  have been reported in Ref. \cite{Ono_INS}; unfortunately,  due to the limited spectral resolution many important details of the magnetic excitation spectrum     appear  missing. Nevertheless, ESR experiments on Cs$_2$CuBr$_4$ in the fully spin-polarized phase above $H_{sat}\sim 30$ T  \cite{CCB_Zvyagin}
detected the exchange mode, allowing to estimate parameters of the effective spin Hamiltonian, $J/k_B = 14.9$ K, $J'/k_B = 6.1$ K   (where $J$ and $J'$ are the exchange coupling parameters along the horizontal and zigzag bonds, respectively, see Fig.\ \ref{fig:CCB_FFD_b}, inset). Besides, the  ESR experiments revealed a substantial zero-field gap, $\Delta\approx 200$ GHz, whose nature has so far remained unclear. In this work, we continue high-field ESR studies of Cs$_{2}$CuBr$_{4}$, evidenced by its remarkable spin dynamics.

Cs$_{2}$CuBr$_{4}$  single crystals  were synthesized by slow evaporation of aqueous solutions of CsBr and CuBr$_2$ similar to the procedure  described in Ref. \cite{Morozin}. Experiments were done using high-field  ESR installations at the Dresden High Magnetic Field Laboratory  (HLD, Germany),  the Center for Advanced High Magnetic Field Science (Osaka University, Japan), and the National High Magnetic Field Laboratory (Florida State University, Tallahassee, USA) \cite{Zvyagin_ESR}. The experiments were done in the frequency range of 100 - 800 GHz, using tunable sources of millimeter-wave radiation sources backward wave oscillators (product of ISTOK, Russia), VDI transmit systems (product of Virginia Diodes Inc., USA), and MVNA vector network analyzer (product of AB Millimetre, France).

A single-line ESR absorption  was detected in the paramagnetic phase ($T\gg J/k_B$), yielding  $g_a=2.15(2)$, $g_b=2.07(2)$, $g_c=2.23(2)$ ($T=294$ K) and $g_a=2.15(2)$, $g_b=2.11(2)$, $g_c=2.26(2)$ ($T=77$ K). Pronounced evolution of the ESR spectrum was observed  upon cooling.

\begin{figure}[h]
\begin{center}
\vspace{-0.5cm}
\includegraphics[width=0.45\textwidth]{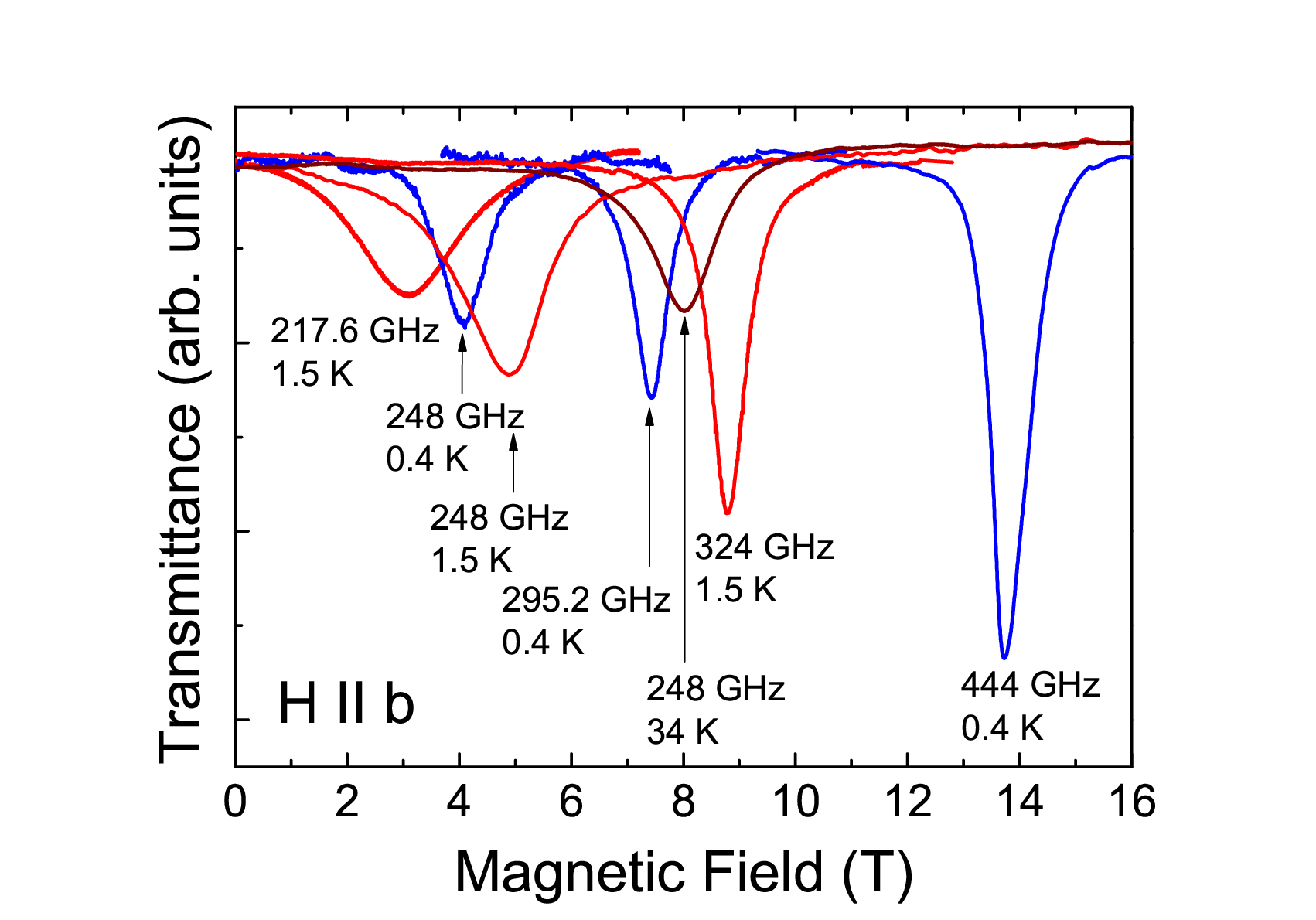}
\vspace{0.0cm}
\caption{\label{fig:CCB_Spectra} (color online) Examples of ESR spectra taken at different frequencies and temperatures with magnetic field applied along the $b$ axis.}
\end{center}
\end{figure}

The frequency-field diagram of  magnetic excitations in Cs$_2$CuBr$_4$ measured  between  0.4 and  4 K  for  magnetic fields, $H$,
applied along the $b$ axis is shown in Fig.\ \ref{fig:CCB_FFD_b}. Some examples of ESR spectra of magnetic excitations at different frequencies are presented in Fig.\ \ref{fig:CCB_Spectra}.
For this orientation, a single gapped mode (labeled B) was observed, whose  frequency-field dependence can be described using the equation:
\begin{equation}
\label{CCB_FFD_b} h\nu = \sqrt{(g^*\mu_B H)^{2}+\Delta^{2}},
\end{equation}
where $\nu$ is the resonance frequency, $h$ is the Planck constant, $\mu_B$ is the Bohr magneton, $g^*$ is the paramagnetic $g$ factor  (the $g$ factor measured at $T=77$ K, $g^*=2.11$, was used for the fit and calculations), and  $\Delta$ is the zero-field excitation  gap. The best fit was obtained using $\Delta=199(4)$ GHz (which corresponds to 9.5 K) for  $T=1.5$ K.

\begin{figure}[h]
\begin{center}
\vspace{-0.0cm}
\includegraphics[width=0.47\textwidth]{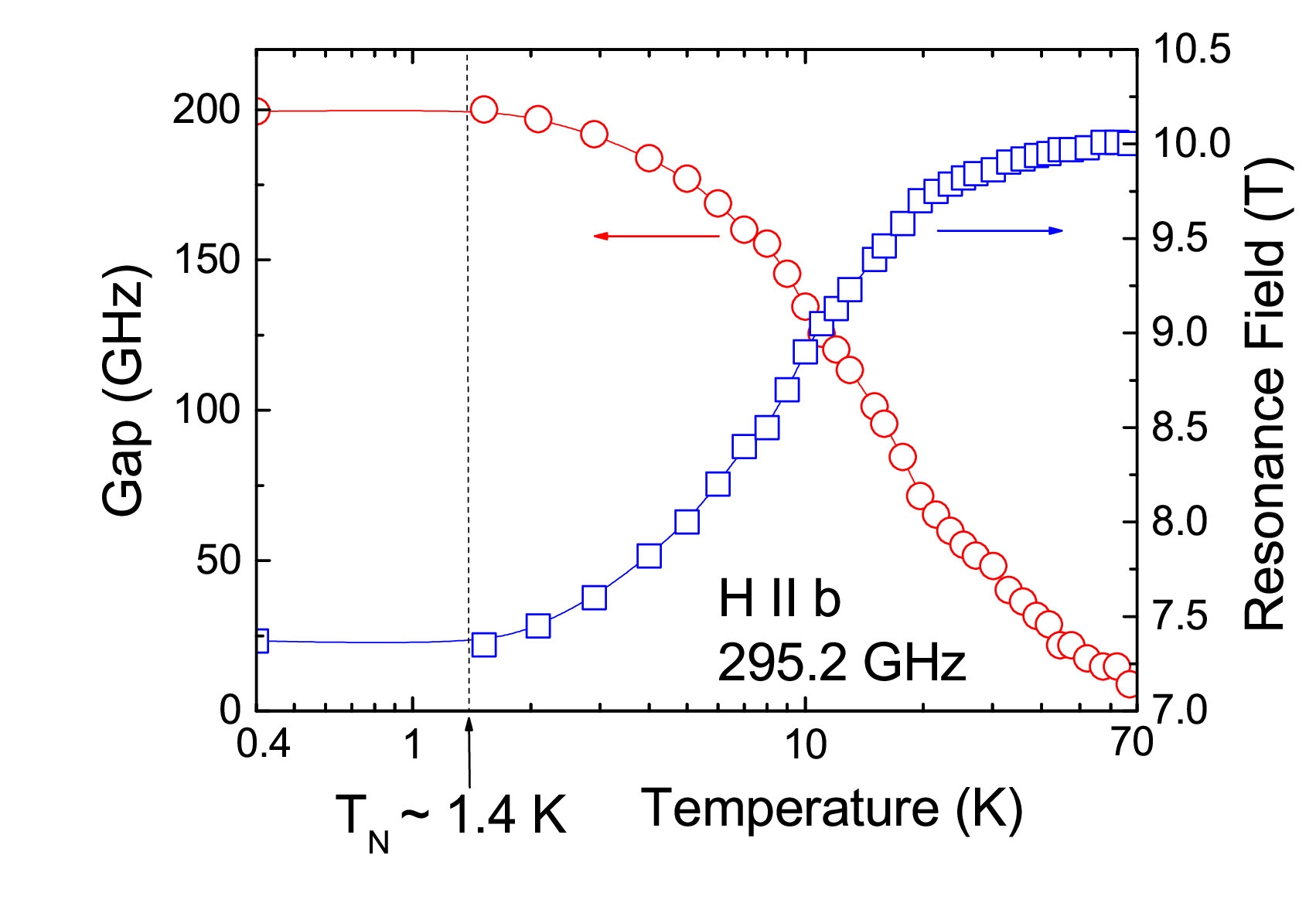}
\vspace{0.0cm}
\caption{\label{fig:CCB_TD} (color online) Temperature dependence of the resonance field  (squares; the data are taken at  295.2 GHz with magnetic
field applied along the $b$ axis) and the zero-field gap size (circles) calculated using Eq. (\ref{CCB_FFD_b}) (see the text for details).  Lines  are guides for the eye.
}
\end{center}
\end{figure}

The temperature dependence of the ESR field position at a frequency of 295.2 GHz   with magnetic field applied along the $b$ axis
is shown in Fig.\ \ref{fig:CCB_TD} by  squares. Using these data and Eq. (\ref{CCB_FFD_b}),    the gap size $\Delta$ can be calculated for different temperatures   (circles in Fig.\ \ref{fig:CCB_TD}).  We found that the gap in Cs$_2$CuBr$_4$  persists up to relatively high temperatures ($T \sim J/k_B$ and above), suggesting the presence of short-range-order spin correlations, responsible for the gap opening,  well above $T_N$.

\begin{figure}[h]
\begin{center}
\vspace{-0.0cm}
\includegraphics[width=0.5\textwidth]{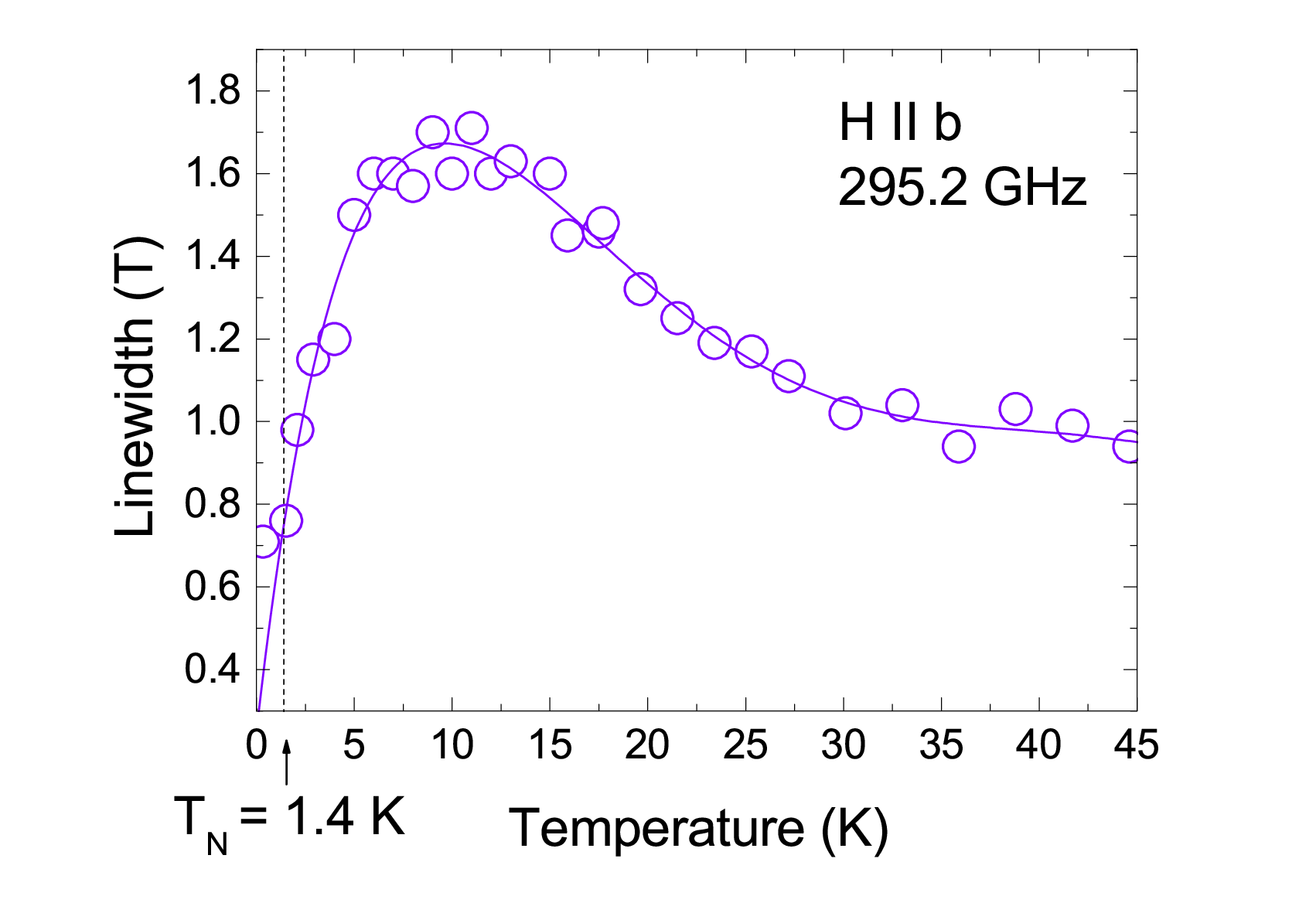}
\vspace{0.0cm}
\caption{\label{fig:CCB_LW_TD} (color online) Temperature dependence of the ESR linewidth measured  at a frequency of 295.2 GHz with magnetic
field applied along the $b$ axis.   Line is guide for the eye.
}
\end{center}
\end{figure}

\begin{figure}[h]
\begin{center}
\vspace{0.0cm}
\includegraphics[width=0.5\textwidth]{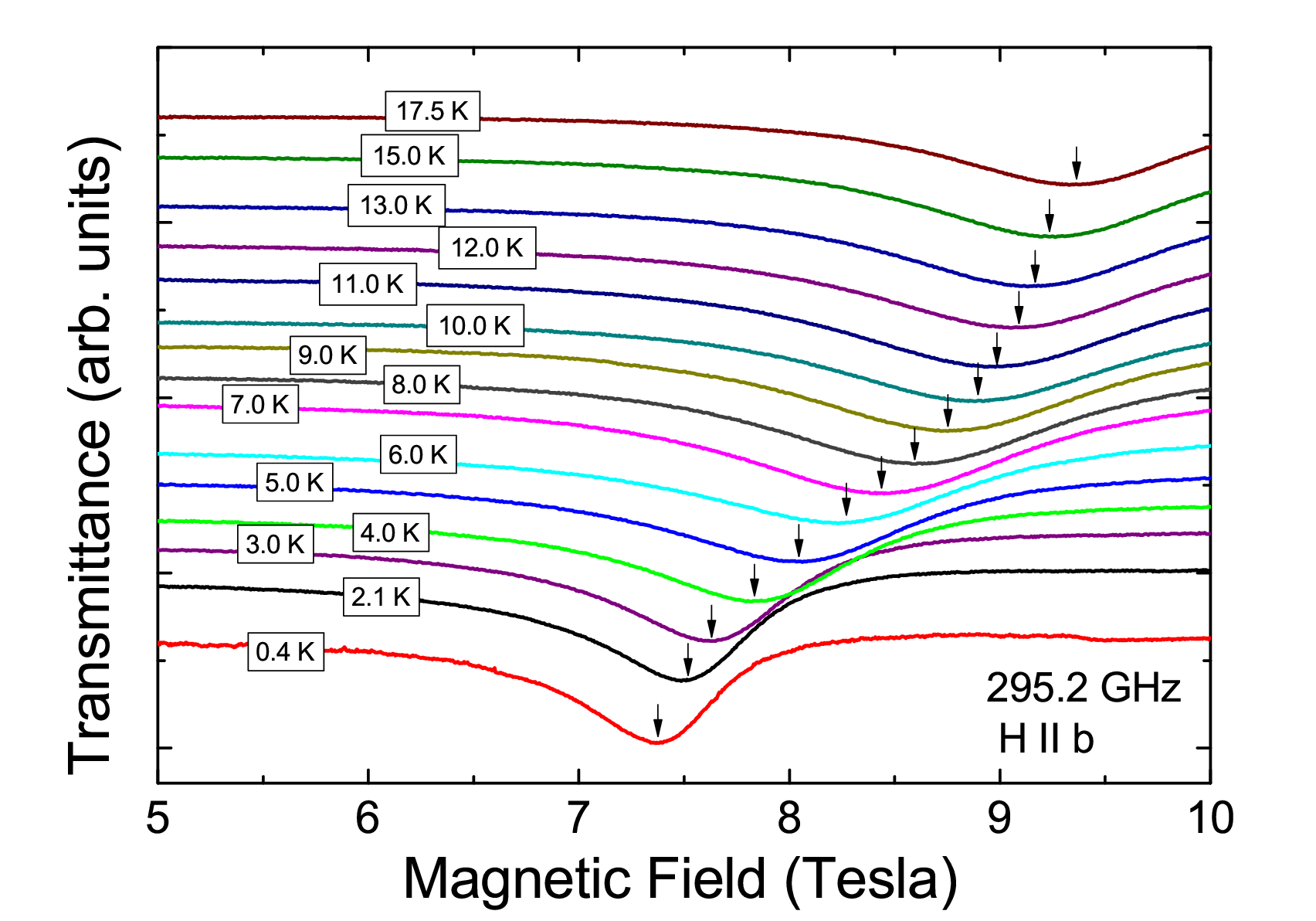}
\vspace{0.0cm}
\caption{\label{fig:CCB_SPECTRA_b_T} (color online) Temperature evolution of ESR spectra  taken at  295.2 GHz with magnetic field applied along the $b$ axis (mode B).}
\end{center}
\end{figure}

At  $T_N=1.4$ K Cs$_2$CuBr$_4$ undergoes a transition into a 3D long-range ordered phase.  Noticeably, the transition   leaves the gap size almost unchanged  (Fig.\ \ref{fig:CCB_TD}).  Similar effects  were observed   in   a number of quantum  AFs \cite{Zal,Coldea_coex,Lake1,Lake2},  whose ground states below $T_N$ are 3D magnetically ordered and the low-energy excitation spectra are  determined by 3D long-range-order correlations, while the high-energy ($\sim J$) spin dynamics  is still determined by low-D effects. Remarkably, no ESR line-width anomaly  was observed in Cs$_2$CuBr$_4$ at  the transition into the 3D magnetically ordered state (Fig.\ \ref{fig:CCB_LW_TD}). The temperature evolution of ESR spectra  taken at  295.2 GHz with magnetic field applied along the $b$ axis  is shown in Fig.\ \ref{fig:CCB_SPECTRA_b_T}.

Overall, the  ESR properties of Cs$_2$CuBr$_4$  in many respects appear very similar to those obtained for Cs$_2$CuCl$_4$ \cite{Povarov}. On the other hand, contrary to Cs$_2$CuCl$_4$ with the zero-field energy gap $\Delta \sim D \approx 0.15 J$, the  size of the gap in  Cs$_2$CuBr$_4$ ($\Delta \approx 0.6J$) is too large to be explained in terms of the quasi-1D Heisenberg AF chain model with the uniform  DM interaction $D$.  This strongly suggests that,   compared to Cs$_2$CuCl$_4$,  the effect of frustrated interactions  on the spin dynamics (and more generally, on the  magnetic properties) in Cs$_2$CuBr$_4$ can be  different.

To get an insight into the spin dynamics of Cs$_2$CuBr$_4$, we did simple spin-wave calculations
for the ordered state of an orthorhombically-distorted triangular-lattice antiferromagnet.
The minimal spin Hamiltonian includes only exchange interactions
\begin{equation}
\hat{\cal H} = \sum_{\langle ij\rangle} J_{ij}\, {\bf S}_i\cdot {\bf S}_j \,,
\label{H0}
\end{equation}
where the nearest-neighbor exchange constant is  $J_{ij}=J$ or $J'$ for horizontal or zigzag bonds, respectively
(see the inset in Fig.~\ref{fig:CCB_FFD_b}) \cite{Tanaka02,rem3}. The classical ground state of the spin model (\ref{H0}) is a planar spiral magnetic structure with  the ordering wave-vector
${\bf Q}=(0,Q,0)$ and  $\cos(Qb/2)=-J'/2J$.  At zero temperature, the energy of magnetic excitations in a spiral AF
is given by the standard expression \cite{Veillete05,Dalidovich06}:
\begin{equation}
\epsilon_{\bf k} = S\sqrt{(J_{\bf k} - J_{\bf Q})
({\textstyle \frac{1}{2}}J_{\bf k+Q}+{\textstyle \frac{1}{2}}J_{\bf k-Q}-J_{\bf Q})} \ ,
\label{Ek}
\end{equation}
where $J_{\bf k} = \sum_j J_{ij} e^{i{\bf k}({\bf r}_{i}-{\bf r}_{j})}$ is the Fourier transform of the exchange interactions.

\begin{figure}
\begin{center}
\vspace{0.0cm}
\hspace{-0.0cm}
\includegraphics[width=1\columnwidth]{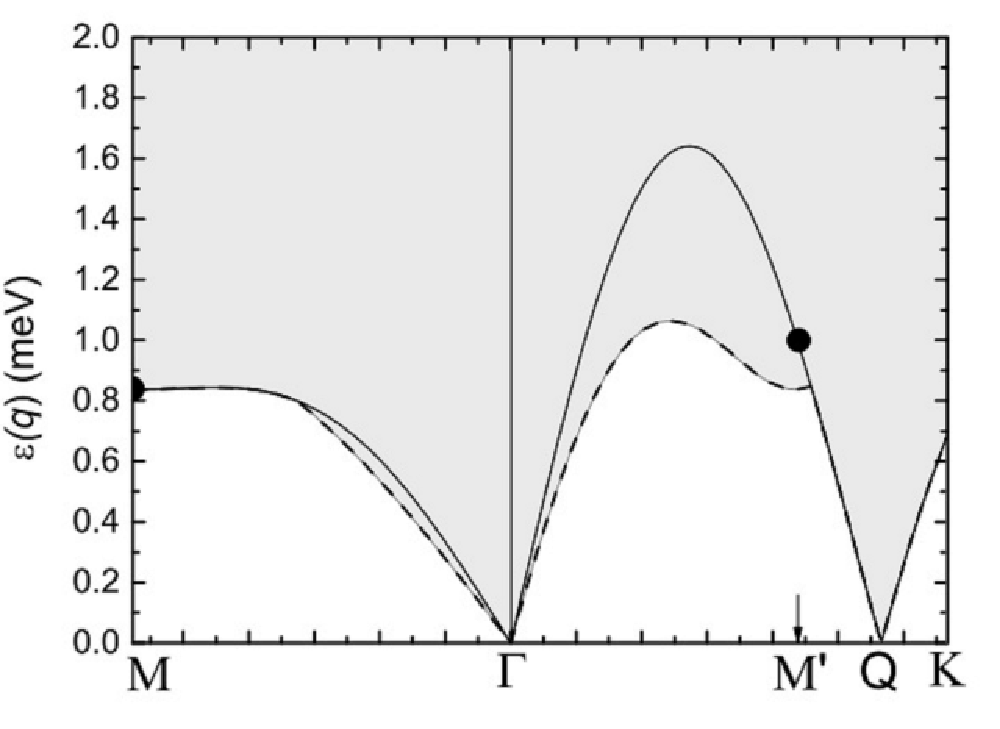}
\vspace{0cm}
\caption{\label{fig:disp}  Calculated magnon dispersion in Cs$_2$CuBr$_4$.
The $\Gamma$ point refers to the center of the Brillouin zone, whereas Q is the ordering wave vector.
The rf field in ESR experiments can excite magnons at points M and M$'$ due to the staggered DM interaction.
The two-magnon continuum is shown in gray color.
}

\end{center}
\end{figure}

In the framework of the formulated toy model, the origin of the gap may be understood if real geometry of
DM interactions in Cs$_2$CuBr$_4$ is taken into account \cite{Starykh,CCB_Zvyagin}.
The staggered component of the DM interaction reduces the translational symmetry of the spin Hamiltonian (\ref{H0})
producing a doubling of the unit cell in the $bc$ plane and, consequently, folding of the Brillouin zone.
As a consequence, an excitation with a momentum $\bf k$ mixes with another magnon at
$\bf k + q_c^*$, where ${\bf q_c}^*=(0,0,2\pi/c)$ is a new reciprocal lattice vector.
The actual shift of the magnon energy (\ref{Ek}) produced by the DM interactions may be quite small
because of smalness of $D/J$. However, the presence of a nonzero mixing matrix element
leads to a finite ESR response for the high-energy magnon at the M point (${\bf q_c}^*$) in the unfolded Brillouin zone, see Fig.~\ref{fig:disp}.
It is worthwhile to mention that ESR transitions at the Brillouin zone boundary were observed   in a number of  low-D quantum magnets (see, e.g., Ref. \cite{CuGeO3,Hald,DK,Wang}), while the corresponding selection rules  were theoretically studied in Ref. \cite{Sakai}. Remarkably, as  revealed  experimentally \cite{DK,Wang},   such nominally  forbidden ESR transitions (which become allowed in the presence of the DM interaction)
 can be very intensive,  exhibiting, on the other hand,   a very pronounced polarization dependence.

The corresponding ESR gap is calculated using Eq.~(\ref{Ek}) as
\begin{equation}
\Delta=\epsilon_{\bf q_c^*} = 4J'S\Bigl(1-\frac{J'}{2J}\Bigr)\,.
\label{Delta12}
\end{equation}
With  $J/k_B=14.9$~K and $J'/k_B=6.1$~K determined in the high-field ESR experiments \cite{CCB_Zvyagin},
the gap for Cs$_2$CuBr$_4$ is estimated to be $\Delta$ = 0.84 meV, which corresponds to 9.7~K or 203 GHz.
This value perfectly agrees with our experimental  observations.
This agreement between the harmonic spin-wave theory for the exchange
model and the experimental results  might be somewhat fortuitous, as we completely neglected
quantum fluctuations and the DM interactions. The observed agreement  may indicate that these additional
effects partially compensate each other \cite{rem2}.

The above consideration also predicts one more ESR mode arising from excitation of magnons with momenta
${\bf k} = \pm{\bf Q}+{\bf q}_b^*$ = $(0,2\pi/b\pm Q,0)$ (M$'$ point in Fig.~\ref{fig:disp}).  The corresponding energy is
\begin{equation}
\Delta'=\epsilon_{{\bf q}_b^* \pm {\bf Q}} =
4J'S\Bigl(1-\frac{J'}{2J}\Bigr)\,\sqrt{1+\frac{J'}{J}+\frac{J'^2}{2J^2}}\,.
\label{Delta23}
\end{equation}
Using the above  $J$ and $J'$ \cite{CCB_Zvyagin} we find $\Delta' \approx 1$~meV.
As further calculations show, these excitations appear well inside the two-magnon continuum (Fig.~\ref{fig:disp}), thus having  a finite lifetime even at zero temperature. Therefore, they cannot be observed in ESR experiments.
This conclusion is in agreement with the existing theoretical results for the damping of high-energy magnons in the spin-1/2
Heisenberg AF on a perfect triangular lattice \cite{Rev,Chern09,Mou}.

\begin{figure}
\begin{center}
\vspace{-0.5cm}
\includegraphics[width=0.53\textwidth]{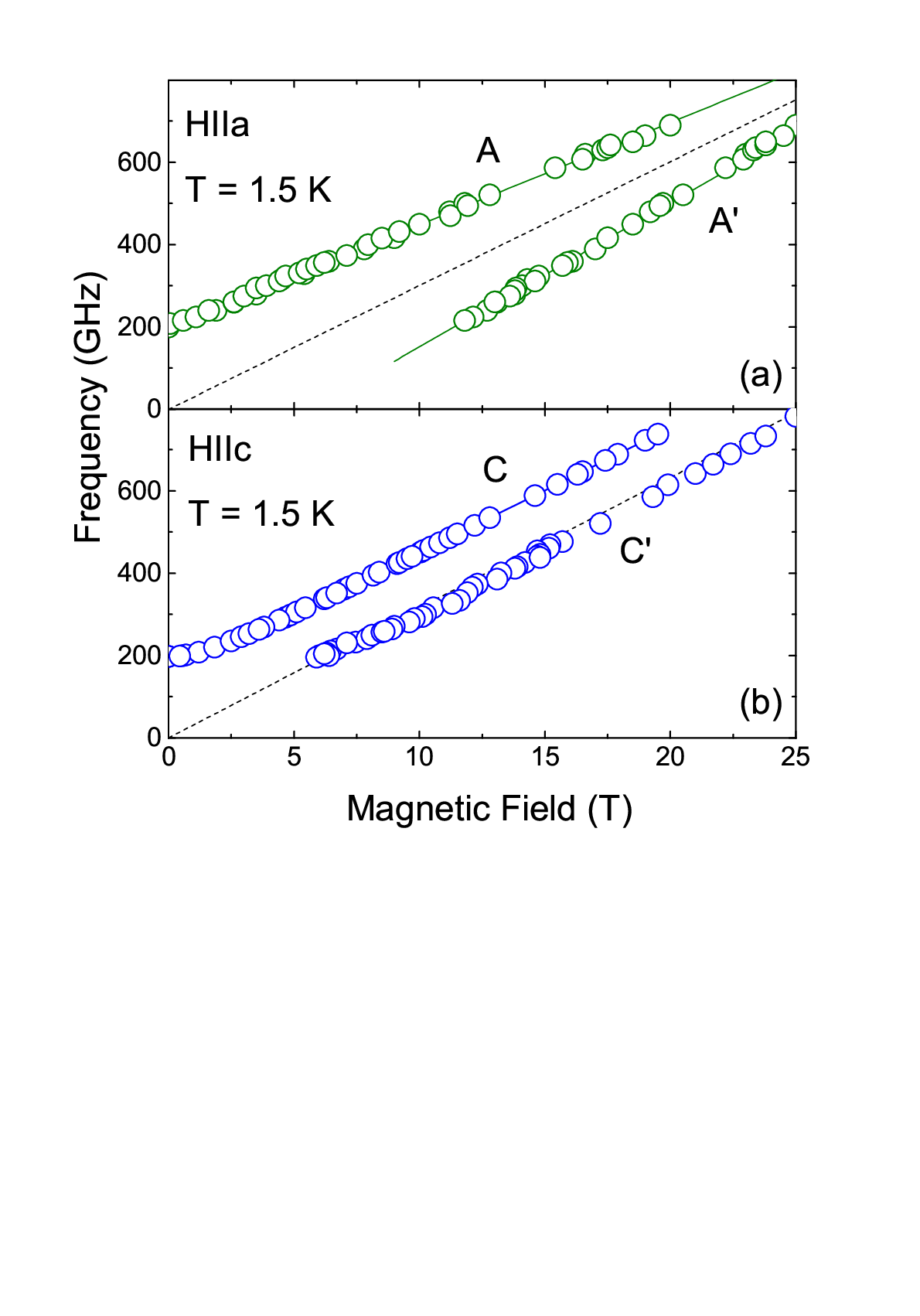}
\vspace{-5cm}
\caption{\label{fig:CCB_FFD_ac} (color online) Frequency-field diagrams of ESR excitations
with magnetic field applied along $a$ (a) and $c$ (b) axes.  The dashed lines correspond to  paramagnetic resonances with $g_a=2.15$ ($H \|a$,  $T=77$ K) and $g_c=2.26$ ($H \|c$,  $T=77$ K).  Solid lines  are guides for the eye. Insets show  examples of the corresponding ESR spectra. The data are obtained at 1.5 K. }
\end{center}
\end{figure}

\begin{figure}
\begin{center}
\vspace{-0.5cm}
\includegraphics[width=0.53\textwidth]{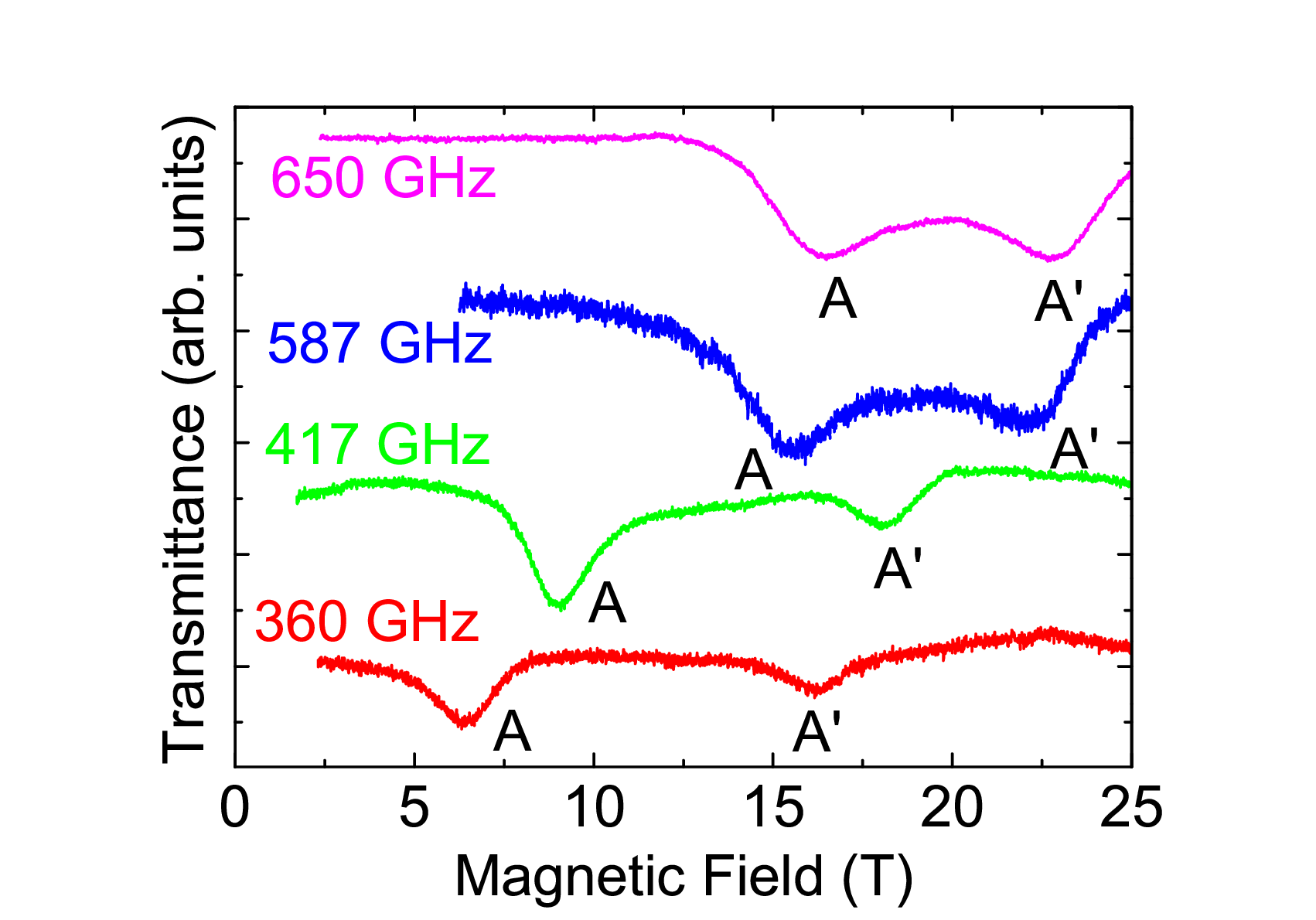}
\vspace{-0cm}
\caption{\label{fig:CCB_Spectra_a} (color online)  Examples of the corresponding ESR spectra with magnetic field applied along $a$ axis. The data are obtained at 1.5 K. }
\end{center}
\end{figure}

\begin{figure}
\begin{center}
\vspace{-0.5cm}
\includegraphics[width=0.53\textwidth]{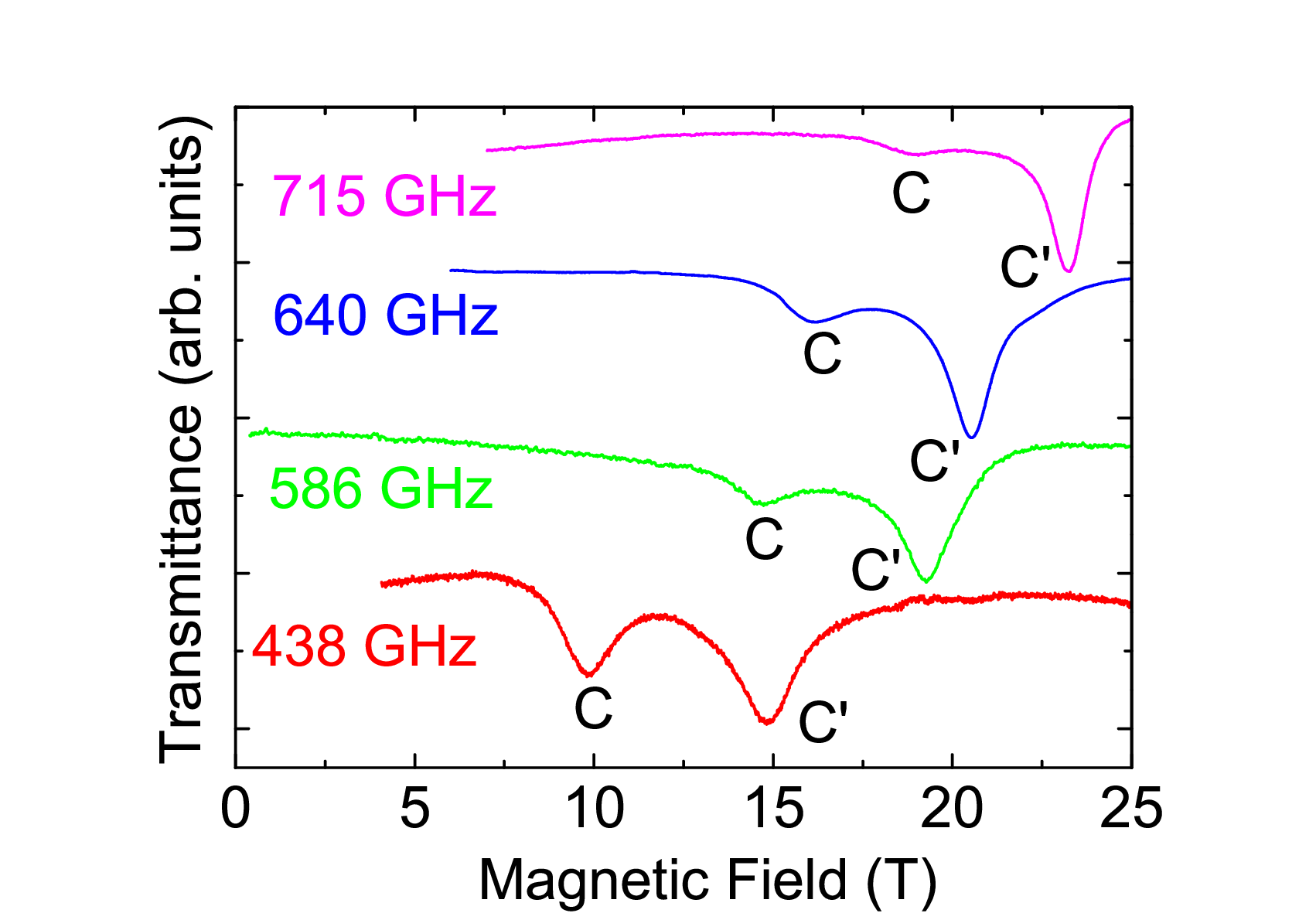}
\vspace{-0cm}
\caption{\label{fig:CCB_Spectra_c} (color online)  Examples of the corresponding ESR spectra with magnetic field applied along $c$ axis. The data are obtained at 1.5 K. }
\end{center}
\end{figure}

For $H \| a$ and $H \| c$, two pairs of ESR modes were observed (A, A$'$, and C, C$'$, respectively); the corresponding
frequency-field  ESR diagrams and ESR spectra  are shown in  Fig.~\ref{fig:CCB_FFD_ac}. The corresponding examples of ESR spectra are shown in  Fig.~\ref{fig:CCB_Spectra_a} and  Fig.~\ref{fig:CCB_Spectra_c}. Upon rotating the sample around  $c$ and $a$ axes,  the   modes A, A$'$ and C, C$'$ (respectively)    merge into the single mode B at $H \| b$, which makes this direction special as it coincides with the direction of the order wavevector $\bf Q$. Noticeably,  approaching the frequency 200 GHz,  modes A$'$ and C  became broader and weaker, so that no  ESR was observed below 200 GHz.

It is important to mention, that due to the development of long-range 3D correlations  below $T_N$ one would expect the emergence of lower-frequency antiferromagnetic resonance (AFMR) magnon modes.   Theory predicts the presence of at most three relativistic Goldstone AFMR modes as a consequence of the complete breaking of the rotational SO(3) symmetry in anisotropic spin systems with a noncollinear ground state \cite{Andreev}. Such  AFMR  modes have been recently observed in the isostructural compound Cs$_2$CuCl$_4$  below $T_N=0.62$ K \cite{Smirn_AFMR}.

In conclusion, our work  provides new insights into the unconventional  spin dynamics  in the spin-1/2 triangular-lattice antiferromagnet
 Cs$_{2}$CuBr$_{4}$, studied  by means of high-field ESR spectroscopy.  In particular,  the temperature behavior of the observed 
zero-field energy gap is investigated in a broad temperature range,  indicating that the gap is a characteristics of short-range-order spin correlations, which appear to persist in  Cs$_{2}$CuBr$_{4}$  down to well below $T_N$. Based on the proposed spin-1/2 distorted triangle-lattice antiferromagnet model,  our findings might suggest,   that Cs$_{2}$CuBr$_{4}$ exhibits a peculiar combination of nearly classical spin dynamics and static magnetic ordering \cite{rem2}  with pronounced quantum effects as revealed, e.g., by the 1/3 magnetization plateau.  This calls for further experimental quantification of the role of quantum fluctuations in Cs$_{2}$CuBr$_{4}$, in particular, for neutron measurements of the ordered magnetic moments and the magnon modes in zero magnetic field.

This work was supported by Deutsche Forschungsgemeinschaft (DFG, Germany). We acknowledge the support of the HLD at HZDR, member of the European Magnetic Field Laboratory (EMFL). A portion of this work was performed at the NHMFL, Tallahassee, FL, which is supported by NSF Cooperative
  Agreement No. DMR-1157490, by the State of Florida, and by the DOE. S.~A.~Z. appreciates the support of the Visiting
  Professor Program at Osaka University. Work at BNL was supported by the U.S. DOE under Contract No. DE-AC02-98CH10886. C.~P. acknowledges the support by the A. von Humboldt Foundation. The authors would like to thank  F.~H.~L. Essler, V.~I. Marchenko, O.~A. Starykh, and A.~I. Smirnov  for fruitful discussions,
   M. Ikeda and T. Fujita for the help in high-field ESR experiments at  AHMF, and S. Miyasaka for his help orienting the samples.


\begin{thebibliography}{10}
\bibitem{rem5}  We have to mention that  in the framework of this interpretation some  questions, such as a relatively small
spin reduction in the 3D magnetically ordered state $\Delta S/S=25\%$ \cite{Spin_Reduction}, remain open.
\bibitem{rem3} It is important to mention that more complex picture of magnetic interactions, including five components of DM interactions (two staggered and three uniform) and a finite interplane  interaction need to be taken into account for  the accurate quantitative description of the spin dynamics in this compound, in particular, in applied magnetic fields.
\bibitem{rem2} Note, that  the ratio $J/J'=0.467$ determined for Cs$_{2}$CuBr$_{4}$  from the experimental value for the zero-field ordering wavevector $\mathbf{Q}=(0,0.575,0)$ \cite{Ono_ENS,Ono_PRB} using the classical expression is very close to $J/J'=0.41(4)$ \cite{CCB_Zvyagin}. This suggests that effects  of quantum fluctuations in Cs$_{2}$CuBr$_{4}$ may be significantly reduced (compared to the isostructural compound   Cs$_{2}$CuCl$_{4}$).
\end{thebibliography}
\end{document}